\pdfoutput=1
\documentclass[journal]{IEEEtran}

\usepackage{cite}

\ifCLASSINFOpdf
   \usepackage[pdftex]{graphicx}
   \graphicspath{{pic/}}
\else
\fi

\usepackage{amsmath}

\usepackage{amsfonts}
\usepackage{derivative} 
\usepackage{physics}
\usepackage{relsize}
\usepackage{subcaption}
\usepackage{authblk}

\begin{document}

\title{Radiation instability of a relativistic electron beam into a split-cavity resonator}

\author{Sergei Anishchenko, Vladimir Baryshevsky, Illia Maroz and Anatoli Rouba}

\affil{Research Institute for Nuclear Problems, Belarusian State University, 11 Bobruiskaya Str., Minsk 220030, Belarus}

\maketitle

\begin{abstract}
The radiation instability in a split-cavity asymmetric resonator is considered for the relativistic case. The space charge of an electron beam is taken into account.
In the small-signal approximation, the energy loss by particles passing through the resonator and the modulation of beam current are investigated.
Based on analytical and numerical calculations, it is shown that the symmetric configuration of a split-cavity resonator provides the highest rate of instability growth. 
It is shown that the beam modulation and the efficiency of energy transfer from particles to electromagnetic field decrease with the increase in the initial electron energy. 
The increase in beam density  has a positive effect on the radiation instability growth. 
It is important to take into account the obtained results when developing generators of electromagnetic radiation and systems for modulating the beam current based on a split-cavity resonator.
\end{abstract}


\IEEEpeerreviewmaketitle

\section{Introduction}

\IEEEPARstart{M}{icrowave} electromagnetic radiation is used in various fields of science and technology, e.g., for plasma heating, radiolocation, acceleration of charged particles, etc. 
Thus, the widespread use of microwave radiation leads to the constantly growing interest in the development and improvement of generators operating in the microwave range.

One direction of the high-power microwave radiation research is devoted to systems based on the transit-time effect. 
Such generators as a transit-time  oscillator (TTO) or monotron~\cite{marcum1946interchange, barroso2000design, barroso2001axial}, split-cavity oscillator (SCO)~\cite{marder1992split}, super-reltron~\cite{miller1994super}, double-foil SCO~\cite{lemke1991theoretical, jun2004new} can produce powerful microwave radiation using high-current electron beams without the need of applying the magnetic field.

Marder and his colleagues presented a detailed computer simulation of particles in the resonator~\cite{marder1992split}. In the case of a small signal, the Marder team developed a method for estimating the resonator parameters providing efficient modulation of a non-relativistic electron beam in the cavity. 

The key element of the generator is a split-cavity resonator. 
This resonator provides the modulation of a high-current electron beam over a short length. 
This makes possible the development of radiation sources without a magnetic guiding system. 
The operation of the Marder generator is based on the transit-time effect arising from the interaction of an electron beam with the electromagnetic field of a split-cavity resonator. 
According to~\cite{marder1992split}, as the result of such interaction, the radiation instability of the electron beam develops. The electron beam becomes modulated. 
It should be emphasized that the structure of the electromagnetic field in the generator allows effective interacting of radiation with a beam, the cross-section of which is proportional to the area of the grid.

However, it should be noted that the abovementioned studies were carried out for the case of non-relativistic beams and symmetrical resonators. 
In this regard, we further consider the case of a relativistic electron beam passing through the asymmetrical resonator. 
The influence of the beam space charge on the modulation efficiency and the rise time of the field in the resonator are also estimated. 
It is shown that a high-current beam can be used to generate stimulated radiation  in the relativistic case.

\section{Interaction of charged particles with an electromagnetic field in resonator}

\begin{figure}[!t]
	\centering
	\includegraphics[width=0.95\linewidth]{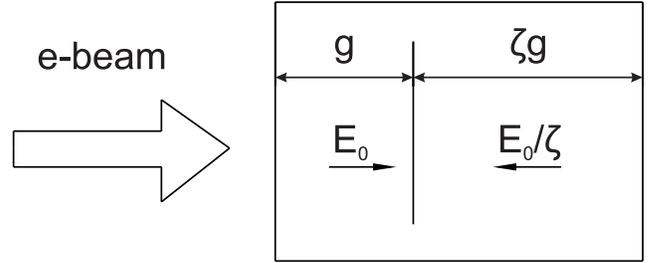}
	\caption{Schematic representation of the asymmetrical resonator.}
	\label{fig:Schematic_resonator}
\end{figure}

\begin{figure}[!t]
	\centering
	\includegraphics[width=0.75\linewidth,height=0.75\linewidth]{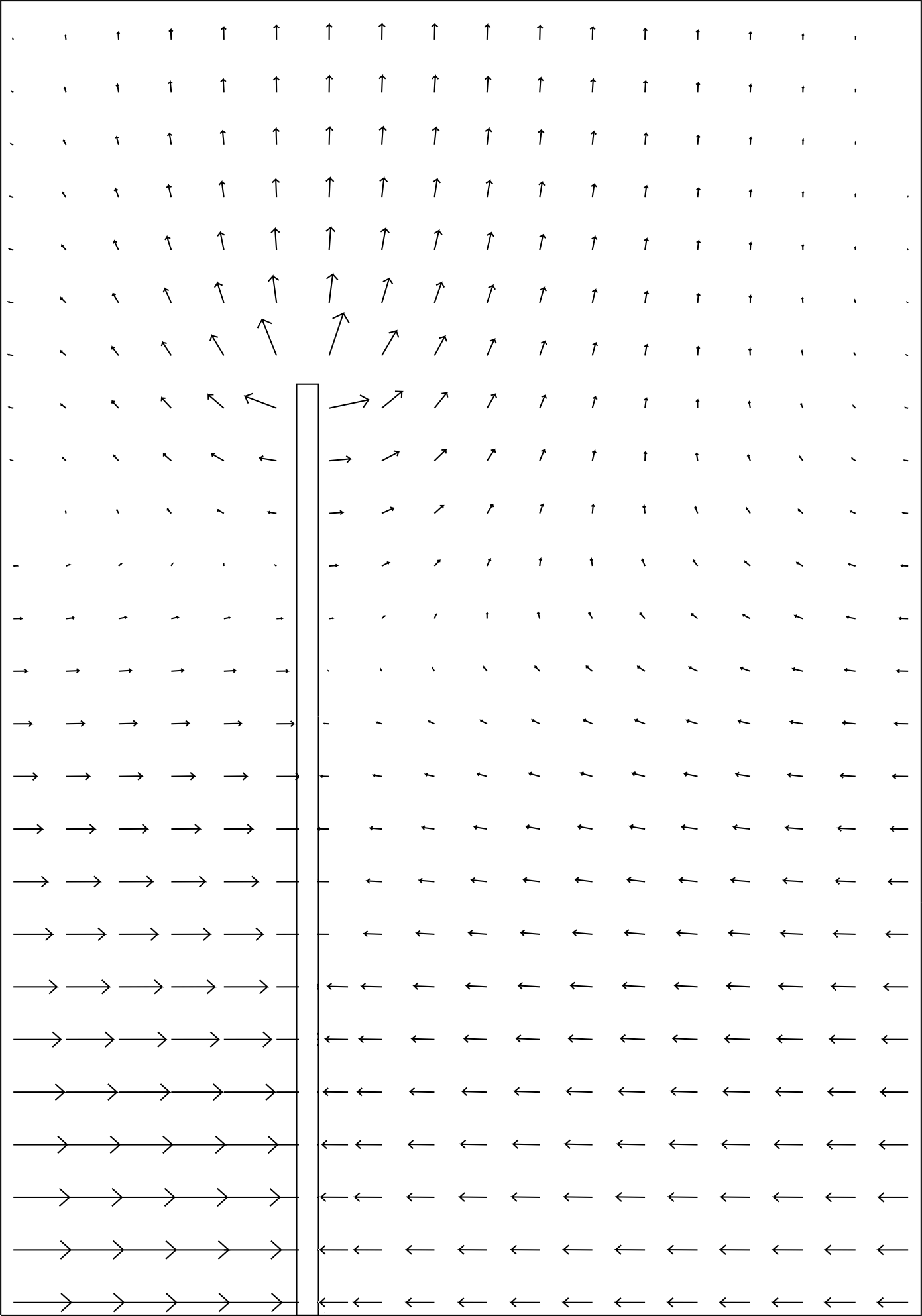}
	\caption{Distribution of the longitudinal component of the electric field of operating mode.}
	\label{fig:Superfish_Ez}
\end{figure}

Let's consider a cylindrical resonator divided into two parts (sections) by a conductive grid or foil transparent for particles (Fig.~\ref{fig:Schematic_resonator}). 
The electrodynamic coupling between the resonator parts is ensured by the gap between the dividing grid and the side wall. 
Such a resonator has an additional set of eigenmodes as compared to a hollow resonator of the same size. 
One of these modes is the operating mode of the generator. 
In the resonator sections, the longitudinal components of the electric field of the operating mode are oppositely directed (Fig.~\ref{fig:Superfish_Ez}).

The electron beam moves along the symmetry axis of the cylindrical resonator and interacts with the operating mode. 
Under the action of the longitudinal component of the electric field, the velocity modulation of the electron beam occurs and the kinetic energy of the particles changes. 
If this change turns out to be negative, then there is an effective transfer of energy from charged particles to the electromagnetic field. As a result, radiation instability arises. 
The emission of electromagnetic waves starts with spontaneous transition radiation.

In the small-signal approximation, the SCO with symmetric resonator was considered in detail in~\cite{marder1992split, baryshevsky2014relativistic}. 
This approximation is applicable for weak fields and, therefore, describes the initial stage of generator operation. 
Now, we investigate the radiation instability of a relativistic beam in a asymmetric resonator. In the case of asymmetric resonator, section lengths are not equal to each other. The small-signal approximation is applied.
The main goal is to determine parameters providing the most efficient energy  transfer from particles to the electromagnetic field. 

For a resonator with unequal sections, the structure of the operating mode was calculated with the Superfish software package~\cite{Poisson_Superfish}. 
Numerical calculations have shown that the longitudinal component of the electric field $E$ acting on a single electron moving near the resonator axis is described by the expression

\begin{equation}\label{eq:E_resonator_definition}
E = 
\begin{cases}
E_0 \sin(\omega t + \Theta) & 0 \leq z < g \\[10pt]
\dfrac{-E_0}{\zeta} \sin(\omega t + \Theta) & g \leq z \leq g (1 + \zeta) 
\end{cases}~,
\end{equation}
where $\Theta$ is the phase of the electromagnetic field at the moment when the particle enters the interaction region, 
$ g $ is the first section length, 
$\zeta$ is the length ratio of two sections, 
$ z $ is the coordinate of the particle, 
$ \omega $ is the operating mode frequency that is specified by the geometry of the resonator, 
and $ E_0 $ is the electric field amplitude. 
Time $ t $ is counted from the moment when the particle enters the resonator. 
One can note that when the particle crosses the separation grid, the electric field changes sign. Also, the absolute value of the field reduces in $ \zeta $ times.

Let's find the change in the kinetic energy of electrons under the action of the field $ E $.
Up to an insignificant factor, the energy change averaged over the phase $ \Theta $ can be written as follows 

\begin{equation}\label{eq:K_normalized_definition_rel}
\Delta \mathbb{ K} = \dfrac{1}{\epsilon^2 x 
	\left( 1 + \dfrac{1}{\zeta} \right) } 
\left\lbrace   \dfrac{1}{2 \pi} \int_{0}^{2 \pi} \left( \dfrac{\sqrt{(\beta \gamma)^2+1}}{\gamma_0} - 1 \right) \mathrm{d}\Theta \right\rbrace~.
\end{equation}
Here, $\gamma = \dfrac{1}{\sqrt{1-\beta^2}}$ is the particle gamma factor at the resonator exit, 
$\beta=v/c$, 
$ v $ is the particle velocity at the resonator exit, 
$ v_0 $ is the initial particle velocity, 
$ c $ is the speed of light in vacuum, 
$\gamma_0$ is the gamma factor of the particles at the resonator entrance,
$\epsilon = \dfrac{e E_0}{m_e c \omega}$ is dimensionless electric field strength,
$ m_e $ is particle mass, $ e $ is elementary charge, $x =  \dfrac{\omega T_0}{2 \pi} $ is the dimensionless length of the first section,
$T_0 = \dfrac{g}{v_0} $ is the time of flight of particles through the first section in the absence of fields. 
It should be noted that the value of $ \epsilon $ is defined in the article differently than in~\cite {marder1992split}.
This is done to leave the dependence of $ \Delta \mathbb{K} $ on the particle velocity $v$ in an explicit form.
In the small-signal approximation, it is assumed that the dimensionless field amplitude $\epsilon$ is small ($\epsilon \ll 1$). 
The normalization on the field energy in the resonator is applied to determine the conditions under which the fastest growth of the modulating field in the resonator is observed. 
With the specified normalization, $\Delta \mathbb{ K}$ turns out to be proportional to the instability increment.

To calculate $\Delta \mathbb{ K}$, it is necessary to find the particle velocities at the resonator exit. 
Therefore, the equations of particles motion in two sections are to be solved. 
For the first section, the equation can be written as

\begin{equation}\label{eq:Eqs_motion_definition_rel}
\dv{p}{t} = - e E_0 \sin(\omega t + \Theta)~,
\end{equation}     
and initial conditions are

\begin{equation}\label{eq:Init_conditions_definition_rel}
\begin{cases}
\left. z \right\vert_{t=0}  = 0 \\[10pt]
\mathlarger {\left. \dv{p}{t} \right\vert_{t=0}} = p_0 
\end{cases}~,
\end{equation}
where $ p $ is the electron momentum at the time $ t $, 
$ p_0 $ is the initial electron momentum. 
Integrating the expression~(\ref{eq:Eqs_motion_definition_rel}) and dividing it by $ c m_e $, one obtains

\begin{equation}\label{eq:Series_section_1_rel}
\beta \gamma =\beta_0 \gamma_0 + \epsilon \left( \cos(\omega t + \Theta) - \cos(\Theta) \right) = F_0(t)~.
\end{equation}

Let's solve the equation~(\ref{eq:Series_section_1_rel}) with respect to the particle velocity $v=\beta c$. 
As the particle moves in the positive $ z $ direction, one gets:

\begin{equation}\label{eq:Solve_defenition_section_1_rel}
z = \int_{0}^{t} \dfrac{c F_0(t')}{\sqrt{1+F_0^2(t')}} \mathrm{d}t'~.
\end{equation}

In the small-signal approximation ($\epsilon \ll 1$), the time of flight of a particle through the first section is given by the expression

\begin{equation}\label{eq:Time_section_1_rel}
T_1 = T_0 + \dfrac{\epsilon \left\lbrace   T_0 \omega \cos(\Theta) + \sin(\Theta) - \sin(T_0 \omega +\Theta)  \right\rbrace  }{\gamma_0^3 \beta \omega}~.
\end{equation}

For the second section of the length $\zeta g$, the equations of motion and initial conditions are similar. 
But the initial velocity of each particle is the velocity $ v_1 $ of the particle when it enters the second section. 
In this case, the phase at the moment of entry is equal to $ (\omega T_1 + \Theta) $. 
As a result, for $\Delta \mathbb{ K}$, one obtains the following expression

\begin{equation}\label{eq:dK_final_rel}
\begin{aligned}
&\Delta \mathbb{ K} = 
 -\dfrac{1}{\zeta L \gamma_0^4}
\left\lbrace -1 - \zeta - \zeta^2 
+ \zeta (1 + \zeta) 	\times \right. \\[10pt]
&\times \cos \left( \dfrac{2 \pi L}{1+\zeta}\right)+ (1 + \zeta) \cos\left( \dfrac{2 \pi \zeta L}{1+\zeta}\right)  
 + \\[10pt] 
 &\left.  + \zeta \bigg[   -\cos(2 \pi L) + 									 
+\pi L \left( \sin\left( \dfrac{2 \pi L}{1+\zeta}\right) + \right.  \right. \\[10pt] 
 &\left. \left. \left. + \sin \left( \dfrac{2 \pi \zeta L}{1+\zeta}\right) 
- \sin(2 \pi L) 
\right)  
\right] 
\right\rbrace  
\end{aligned}~.
\end{equation}
where $ L = x (1 + \zeta) $ is the dimensionless length of the system.

For symmetric resonator, (\ref{eq:dK_final_rel}) is simplified as

\begin{equation}\label{eq:dK_simmetric_final_rel}
\begin{aligned}
\Delta \mathbb{ K} = &
-\dfrac{8}{ L \gamma_0^4}
\left\lbrace \pi L \cos\left( \dfrac{\pi L}{2}\right)  - \sin\left( \dfrac{\pi L}{2}\right)   \right\rbrace \times \\[10pt]
& \times \sin^3 \left( \dfrac{\pi L}{2}\right)
\end{aligned}~.
\end{equation}

Expression~(\ref{eq:dK_simmetric_final_rel}) coincides with that given in~\cite{marder1992split} up to a coefficient $\dfrac{1}{ 2 \gamma_0^4}$. 
The $1/2$ appears due to the differences in the normalization and definition of $\epsilon$, and the factor $\gamma_0^{-4}$ arises due to relativity. 
The presence of the coefficient $\dfrac{1}{ 2 \gamma_0^4}$ in expressions~(\ref{eq:dK_final_rel}) and~(\ref{eq:dK_simmetric_final_rel}) indicates a decrease in energy losses with an increase in the Lorentz factor $\gamma_0$. 
Consequently, the Marder's generator is suitable for operation with weakly and moderately relativistic beams.

\begin{figure}[!t] 
	\centering
	\includegraphics[width=0.75\linewidth]{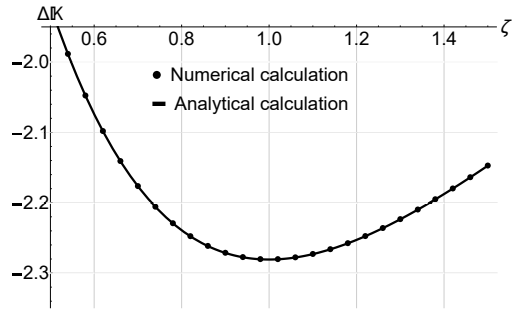}		
	\caption{Dependence of the electron energy losses on the ratio of the section lengths $\zeta$ at $ L = 0.53 $.}
	\label{fig:dK_k__rel}
\end{figure}

\begin{figure}[!t] 
	\centering
	\includegraphics[width=0.75\linewidth]{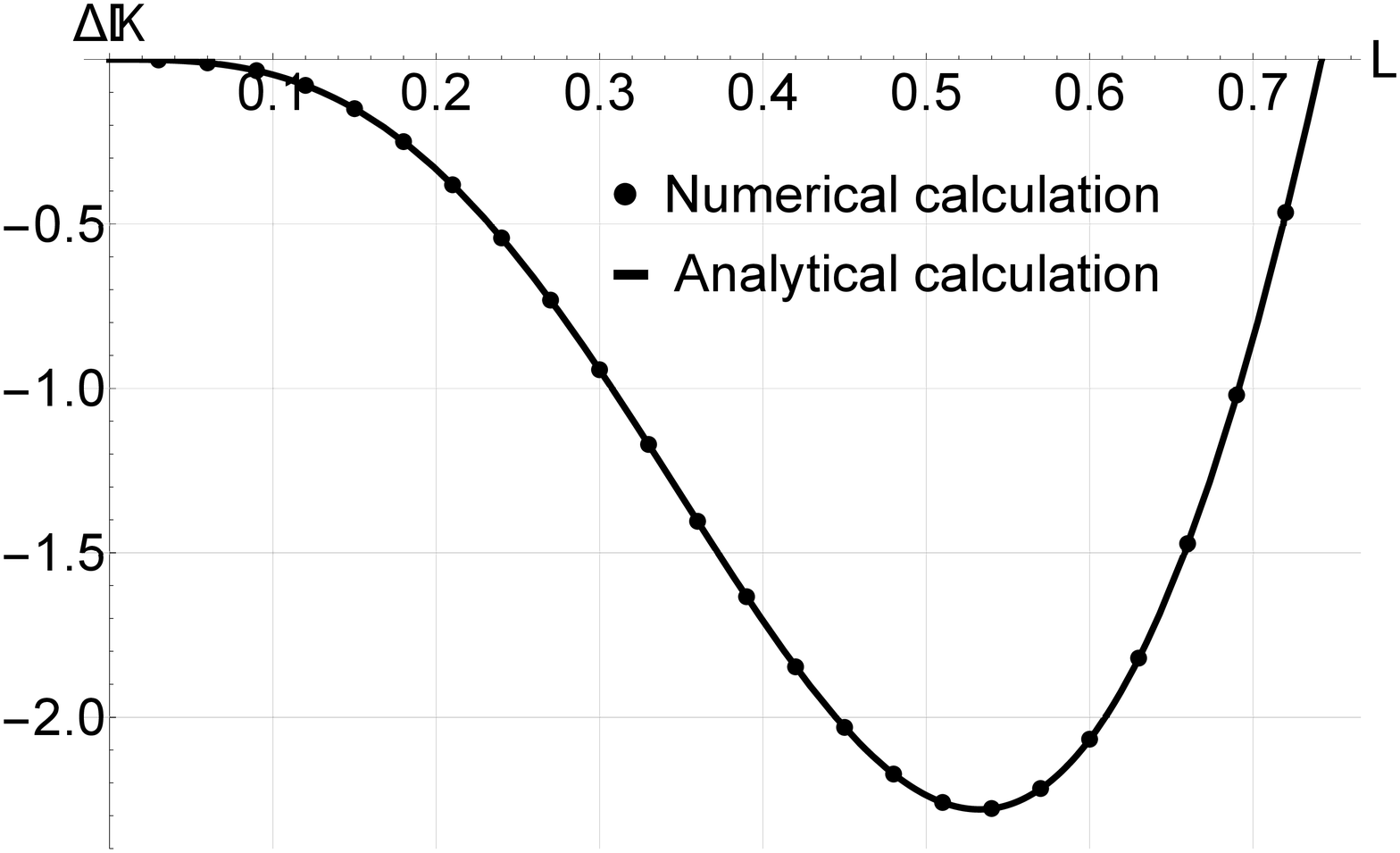}	
	\caption{Dependence of the electron energy losses on the cavity size $ L $ at $ \zeta = 1 $.}
	\label{fig:dK_L__rel}
\end{figure}

Without loss of generality, further calculations is performed at a fixed value $ \beta_0 = 0.1 $. 
As follows from the dependence $\Delta \mathbb{ K}$ on $ L $ and $ \zeta $ shown in Fig.~\ref{fig:dK_L,k__rel}, there are such values of parameters $L_0 $ and $ \zeta_0 $ at which the energy losses are maximal: $ L_0 = 0.53 $ and $ \zeta_0 = 1 $. 
One notes that the condition $ \zeta_0 = 1 $ corresponds to a resonator with equal section lengths. 
With a fixed cavity length $ L_0 = 0.53 $, the shift of the separation grid away from the symmetric position (Fig.~\ref{fig:dK_k__rel}) leads to insignificant changes in $\Delta \mathbb{ K}$. 
For symmetric configurations ($ \zeta = 1 $), a significant decrease of $\Delta \mathbb{ K}$ is observed when the length $ L $ deviates from the optimal value $ L_0 = 0.53 $ (Fig.~\ref{fig:dK_L__rel}).

\begin{figure}[!t] 
	\centering
	\includegraphics[width=0.75\linewidth]{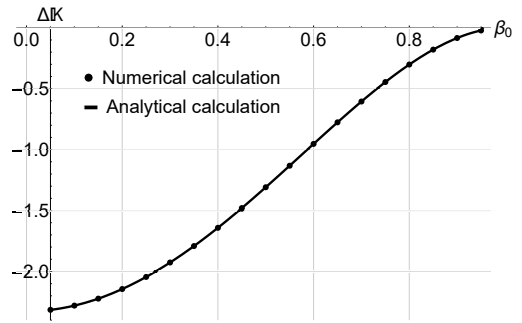}
	\caption{Dependence of the electron energy losses on the initial electron velocity $ \beta_0 $ at $ L= 0.53 $ and $ \zeta = 1 $.}
	\label{fig:dK_beta__rel}
\end{figure}

Energy losses by particles are observed in a wide range of dimensional lengths $ L=\omega g(1+\zeta)/2\pi v_0 $. As a result, it is possible to obtain radiation beam instability in a wide range of electron energies for every fixed length $ g $ (Fig.~\ref{fig:dK_beta__rel}).

\begin{figure}[!t] 
	\centering
	\includegraphics[width=0.75\linewidth]{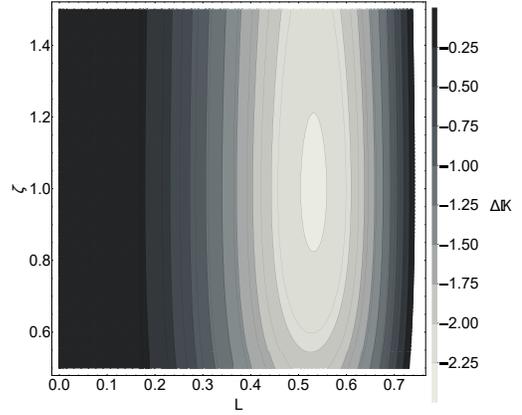}
	\caption{Dependence of the electron energy losses on the resonator size $ L $ and the ratio of the section lengths $ \zeta $.}
	\label{fig:dK_L,k__rel}
\end{figure}

It should be noted that the optimal length of the resonator $ (2 \pi v_0 L_0)/\omega \approx (3.3 v_0)/\omega $ tuned to the specific frequency $ \omega $ decreases with the decrease in initial particle velocity $ v_0 $. 
Consequently, at very low electron energies, the whole length of the system corresponding to the maximum energy losses is small. 
The small length of the resonator leads to the fast overlap of the electrodynamic structure by the plasma formed by the interaction of the beam with foils. As a result, the radiation emission stops. 
Moreover, large energy losses of electrons in the resonator foils is observed  at low particle energies. 
In the ultrarelativistic case $ \gamma_0 \gg 1 $, the energy transferred from the particles to the field decreases noticeably. Besides, the fields required for effective modulation of the electron beam also increase significantly.

In order to obtain higher output power with a specific resonator configuration, it is necessary to increase the beam current. 
This leads to an increase in the current density and, consequently, the space charge field. 
For this reason, it is required to take into account the space charge of electron beam when studying the energy losses of particles.

Let assume that the beam cross-section is uniform and only the longitudinal component of the beam electric field is nonzero. 
In this case, the equation of motion and initial conditions have the following forms, respectively,

\begin{equation}\label{eq:Eqs_motion_1_definition_rel}
\dv{p}{t} = - e E_0 \sin(\omega t + \Theta) + m_e \omega_p^2 \left( z - \dfrac{g}{2} \right)
\end{equation} 
and
\begin{equation}\label{eq:Init_conditions_1_definition_rel}
\begin{cases}
\left. z \right\vert_{t=0}  = 0 \\[10pt]
\mathlarger { \left. \dv{p}{t} \right\vert_{t=0} }= p_0 
\end{cases}~.
\end{equation}
Here, $ \omega_p = \sqrt{\dfrac{n_e e^2}{m_e \epsilon_0}} $ is the plasma frequency, $ n_e $ is the electron concentration, and $ \epsilon_0 $ is the electric constant. 
For further analysis, one defines a dimensionless quantity $S=\dfrac{\omega_p}{\omega}$ characterizing the current density in the system. 
Based on the results above, one finds the optimal values of the system parameters $ (L, \zeta, S) $ providing the maximum energy loss by particles. 
The study is carried out near the values $ L_0 = 0.53 $ and $ \zeta_0 = 1 $. 
The results of the numerical solution of the equation~(\ref{eq:Eqs_motion_1_definition_rel}) are shown in Fig.~\ref{fig:dK_S,k__rel}, Fig.~\ref{fig:dK_S,L__rel} and Fig.~\ref{fig:dK_S,beta__rel}. 
At small values of $S$, the parameters $ L $ and $\zeta$ corresponding to the greatest energy losses practically do not change. 
Thus, at $ S = 0.3 $ and $ S = 0.5 $ the dimensionless lengths are $ L_{0.3} = 0.52 $ and $ L_{0.5} = 0.50 $, respectively. 
The change in the parameter $ \zeta $ is insignificant, and $ \zeta = 1 $ can be assumed. 
The dependence of the $\Delta \mathbb{ K}$ on the particle velocity considering the space charge is the same as without it (Fig.~\ref{fig:dK_S,beta__rel}). 
Thus, the presence of the field of the beam does not lead to a significant change in the optimal parameters of the resonator.
Moreover, the space charge increases the efficiency of energy transfer from particles to the electromagnetic field. 
As shown in~\cite{Maroz2019BSU}, this behavior is also typical for the nonrelativistic case.

\begin{figure}[!t] 
	\centering
	\includegraphics[width=0.75\linewidth]{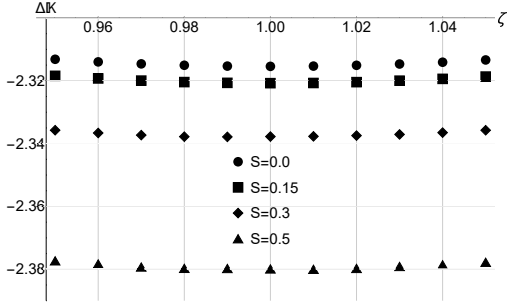}
	\caption{Dependence of the electron energy losses on the ratio of the section lengths $\zeta$ at $ L = 0.53 $ taking into account the space charge.}
	\label{fig:dK_S,k__rel}
\end{figure}

\begin{figure}[!t]
	\centering
	\includegraphics[width=0.75\linewidth]{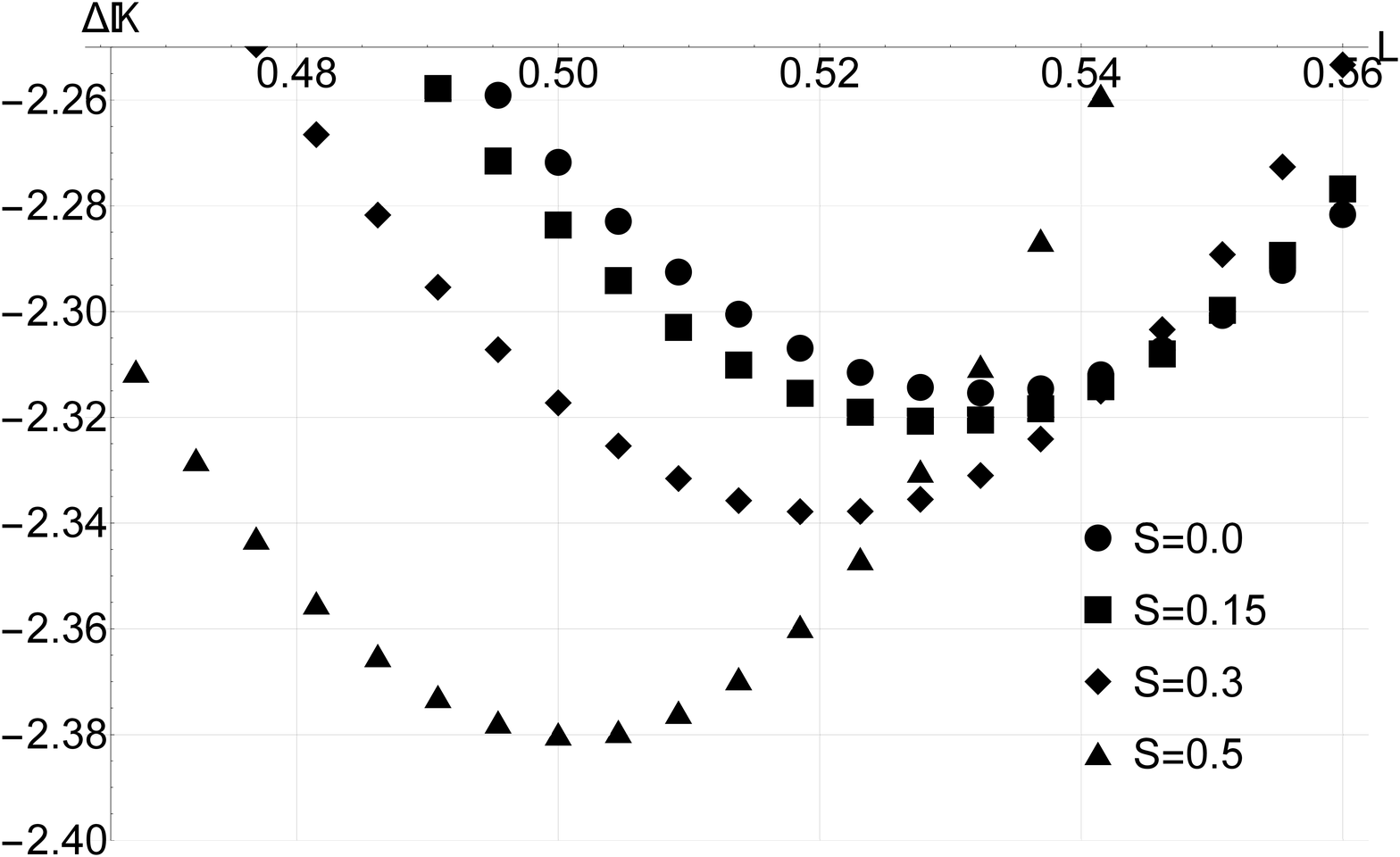}
	\caption{Dependence of the electron energy losses on the cavity size $ L $ at $ \zeta = 1 $ taking into account the space charge.}
	\label{fig:dK_S,L__rel}
\end{figure}

\begin{figure}[!t] 
	\centering
	\includegraphics[width=0.75\linewidth]{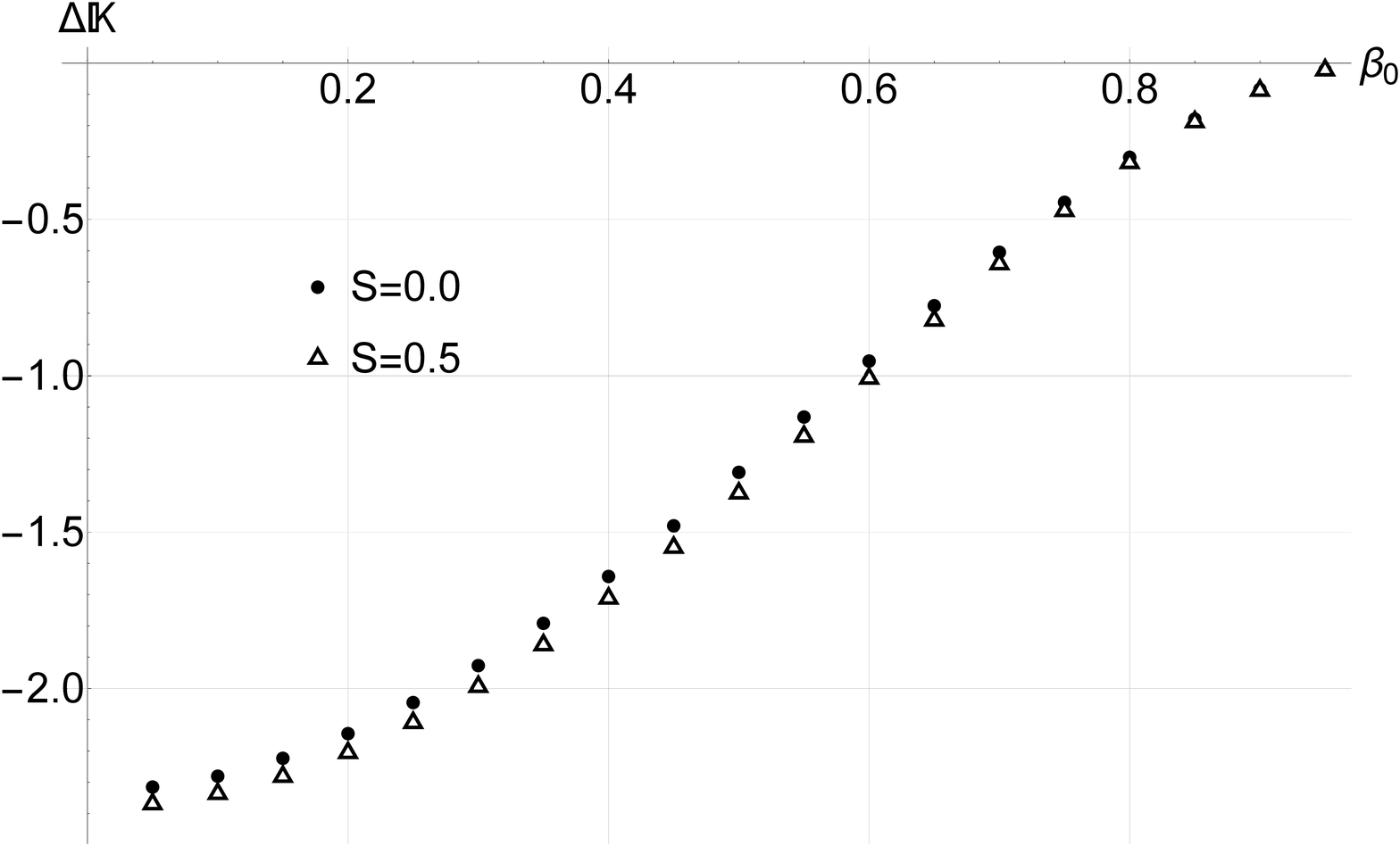}
	\caption{Dependence of the electron energy losses on the initial velocity $\beta_0$ at $ \zeta = 1 $ taking into account the space charge. 
	For each value of $ S $, the dimensionless length $ L $ is chosen to maximize energy losses.}
	\label{fig:dK_S,beta__rel}
\end{figure}

\section{Estimation of the field rise time in the resonator}

The results obtained above allow one to estimate the growth rate of the electromagnetic field in the resonator. 
The initial field in the resonator arises due to the generation of spontaneous transition radiation by the electron beam. 
Spontaneous emission processes determine the field in the cavity only at the initial stage of SCO operation. 
With the instability development, the induced radiation begins to dominate.

Let's estimate the growth rate of the field in the resonator in the small-signal approximation. 
It can be assumed that the distribution of the radiation field in the resonator with a beam is the same as in the resonator without a beam~\cite{jackson1999classical}. 
Then, the field energy in the entire resonator is found by integrating the electromagnetic field density over the resonator volume:

\begin{equation}\label{eq:W_res_definition}
W_{res}= E_0^2 \epsilon_0 \left( \dfrac{V_{res}}{\zeta} \right) \int_{0}^{1} x \mathrm{J_0^2} \left(x \eta_{01} \right) \mathrm{d} x  ~, 
\end{equation}
where $ W_{res} $ is the energy of the electromagnetic field in the resonator, 
$\mathrm{J_0}$ is zero-order Bessel function, 
$ \eta_{01} $ is the first root of the zero-order Bessel function, 
$ E_0 $ is the maximal field amplitude in the resonator, 
$ V_{res} $ is the resonator volume which is defined as $ V_{res} = g (1+\zeta) \pi R^2$, 
$ R $ is the resonator radius.

Let's find the power $ P_{SCO} $ of the radiation generated through the transit-time effect when a homogeneous electron beam is passing through the resonator. 
Combining the expression for the distribution of the electric field in the resonator~\cite{jackson1999classical}, the expressions~(\ref{eq:K_normalized_definition_rel}) and~(\ref{eq:W_res_definition}), one obtains

\begin{equation}\label{eq:P_SCO_definition}
P_{SCO}= \dfrac{2 \gamma_0 e I_0 |\Delta \mathbb{ K}| \int_{0}^{\frac{R_b}{R}} x \mathrm{J_0^2} \left(x \eta_{01} \right) \mathrm{d} x }{m_e \omega^2 \epsilon_0 S_{res} \frac{R_b^2}{R^2} \int_{0}^{1} x \mathrm{J_0^2} \left(x \eta_{01} \right) \mathrm{d} x } W_{res} = \alpha_1 W_{res}~, 
\end{equation}
where $ R_b $ is the electron beam radius, 
$ S_{res} $ is resonator cross-sectional area. 
One notes that the expression for the radiation power~[\ref{eq:P_SCO_definition}] includes $ \Delta \mathbb{ K} $ which determines the change in $ P_{SCO} $ when varying parameters $\zeta$, $ S $, $ L $.

Let the resonator have a Q-factor $ Q > \omega/\alpha_1 $. 
Then, the energy balance equation in the resonator has the form:

\begin{equation}\label{eq:dWres/dt__definition}
\dv{W_{res}}{t} = \left( \alpha_1 -\dfrac{\omega}{Q} \right) W_{res} + P_{tr}~, 
\end{equation}
where $ P_{tr} $ is spontaneous transition radiation power. 

Solving this equation with zero initial condition ($ W_{ers} \vert_{t=0} = 0$), one obtains:

\begin{equation}\label{eq:Wres__solution}
W_{res} = \dfrac{P_{tr}}{\alpha_1 -\dfrac{\omega}{Q}} \left( \exp{\left( \alpha_1 -\dfrac{\omega}{Q}\right) t} - 1 \right)~.
\end{equation}

This expression makes it possible to estimate the instability rise time. 
Let's consider a specific example: the resonator radius $ R $ is 60~mm, the beam radius $ R_b $ is 30~mm, particle energy is 300~keV,  frequency $\omega$ is 3~GHz, $ Q = 10^3$. 
With these parameters, the length of the resonator section is $ g = 20 $~mm ($ \zeta = 1 $) that corresponds to the maximal energy loss by the electron beam. 
Let the beam current be $ I_0 = 3 $~kA ($ S = 0.5 $). 
It can be shown that in this case the characteristic power of the spontaneous transition radiation is be about 110~$\mu$W.
With such parameters, the rise time of instability up to the value $\epsilon = 0.05$ which corresponds to the field strength $ E_0 \approx 16 $~kV/cm is about 40~ns. 
The time constant is $ \tau = \left( \alpha_1 -\dfrac{\omega}{Q} \right)^{-1} \approx 2  $~ns.

\section{Modulation of an electron beam passing through a split resonator}

In the resonator considered above, the non-uniformity in the velocity distribution of charged particles occurs, and, due to the finite dimensions of the system, the non-uniformity of the beam density takes place.  
Thus, it is possible to modulate high-current electron beams~\cite{marder1992split}. The SCO can be used as a modulating system for other generators.

Let's consider the process of electron beam bunching. 
From the moment of time $ t_0 $, $ \Delta N = J_0 \Delta t_0 $ particles enter the resonator during the time $ \Delta t_0 $ ($ J_0 $ is the current of the unmodulated beam). 
Let's define $ t_f $ as the moment when particles, entered the resonator at the time instant $ t_0 $, exit it. 
Then the number of particles leaving the interaction region at the time $ t_f $ is equal to $ \Delta N = J \Delta t_f $. 
Dealing with infinitesimal quantities, one obtains the expression for the beam current $ J $ at the resonator exit

\begin{equation}\label{eq:I_definition}
J = J_0 \left(  \odv{ t_f}{ t_0} \right) ^{-1} ~.
\end{equation}

Solving the equations of motion~(\ref{eq:Eqs_motion_definition_rel}) with the initial conditions~(\ref{eq:Init_conditions_definition_rel}) and the initial phase of the field $(\Theta + \omega t_0)$ yields an explicit expression for the beam current J~(\ref{eq:I_definition}).

For convenience, one introduces the normalized variable component of the beam current $ j $

\begin{equation}\label{eq:j_definition}
j = \dfrac{1}{\epsilon} \left(\dfrac{J}{J_0} -1 \right) ~,
\end{equation}
where $\epsilon'$ is defined as

\begin{equation}\label{eq:epsilon'_definition}
\epsilon'= 
\begin{cases}
\mathlarger { \frac{e E_0}{\zeta m_e v_0 \omega} },	& \zeta < 1 \\[10pt]
\mathlarger { \frac{e E_0}{ m_e v_0 \omega} },  		& \zeta \ge 1
\end{cases}~. 
\end{equation}

It should be noted that the choice of such a normalization is due to the fact that there is a limiting value for electric field strength, limited by the electric breakdown. 

The explicit expression for $ j $ can be written as fallows

\begin{equation}\label{eq:j'_definition_rel}
\begin{aligned} 
j'(L,\zeta, t_0)&= 
\frac{1}{\beta_0 \gamma_0^3} \biggr\lbrace-2 \pi  \zeta L \sin \left(\frac{2 \pi  L}{\zeta+1}+\omega t_0\right)-\\[10pt]
&-\zeta \cos (\omega t_0)+\zeta \cos \left(\frac{2 \pi  L}{\zeta+1}+\omega t_0\right)+\\[10pt]
&+\cos \left(\frac{2 \pi  L}{\zeta+1}+\omega t_0\right)+2 \pi  \zeta L \sin (\omega t_0)-\\[10pt]
&-\cos (2 \pi  L+\omega t_0)\biggr\rbrace 
\times
\begin{cases}
1,	& \zeta < 1 \\[10pt]
\dfrac{1}{\zeta},  		& \zeta \ge 1
\end{cases}~. 
\end{aligned} 
\end{equation}

To achieve the maximum efficiency of the energy conversion from the electron beam to the electromagnetic energy, it is necessary to obtain the maximum modulated beam current. 
For this purpose, the influence of system parameters on modulation should be investigated.

Fig.~\ref{fig:j_rel_L,zeta_} shows the dependence of $ j $ on the parameters $\zeta$ and $ L $ at the specific particle energy, (the $ j $ is normalized to the maximum value in the given range). Based on these results, it can be concluded that the maximum beam modulation will be observed for a symmetric resonator $ \zeta = 1 $ with the maximum possible length. (The length is limited by the condition for radiation instability development $ \Delta \mathbb{ K} < 0 $). 
However, in order to ensure the shortest rise time of the field in the resonator, parameters should be close to $ \zeta = 1 $ and $ L = 0.53 $.

\begin{figure}[!t] 
	\centering
	\includegraphics[width=0.75\linewidth]{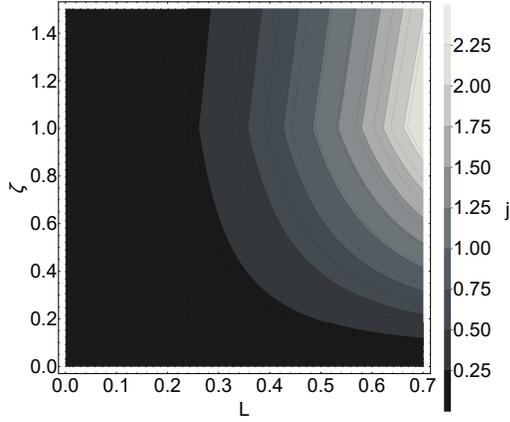}
	\caption{Dependence of $ j $, normalized on the maximum value, on the parameters $ \zeta $ and $ L $.}
	\label{fig:j_rel_L,zeta_}
\end{figure}

As follows from dependence shown in the Fig.~\ref{fig:Num_j(beta,S=0)}, the $ j $ decreases significantly while increasing the particle velocity. 
Consequently, effective current modulation is observed for weakly and moderately relativistic beams.

\begin{figure}[!t] 
	\centering
	\includegraphics[width=0.75\linewidth]{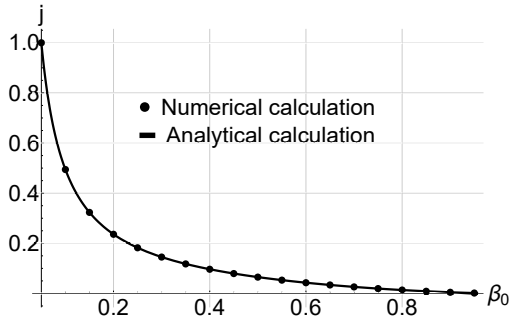}
	\caption{Dependence $ j $ on the initial velocity $\beta_0$ of the particle.}
	\label{fig:Num_j(beta,S=0)}
\end{figure}

Taking into account the space charge of the beam, let's consider the behavior of $ j $. 
Fig.~\ref{fig:j_rel_beta,S_} shows the dependence of the amplitude $ j $ on $ S $ when $ \zeta = 1 $ and $ L = 0.53 $. 
Analysis of Fig.~\ref{fig:j_rel_beta,S_} shows that with an increase in $ S $, the beam modulation increases.

It should be noted that it is necessary to maintain the balance between the energy transferred from the beam to the electromagnetic field and the efficiency of the beam modulation. 
The increase in transfared energy shorten the rise time. The efficient modulation provides the maximal amplitude of the variable component of the beam current. 
While maintaining the balance, it is possible to achieve maximal modulation of the electron beam and radiation emission.

\begin{figure}[!t] 
	\centering
	\includegraphics[width=0.75\linewidth]{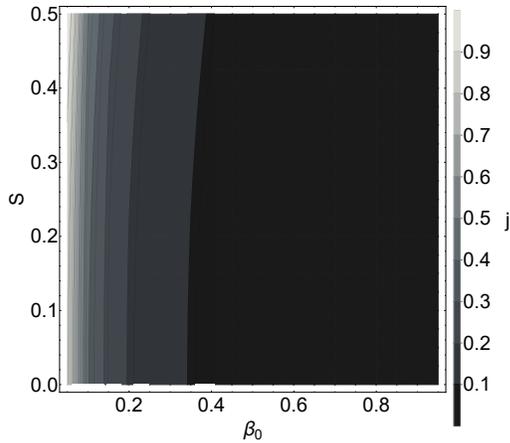}
	\caption{Dependence $ j $ on the initial velocity $\beta$ of the particle.}
	\label{fig:j_rel_beta,S_}
\end{figure}

\section{Conclusions}
In this work, the radiation instability of a relativistic electron beam in the asymmetric split resonator is theoretically investigated. 
In the small-signal approximation, expressions for the energy losses of electrons and the rise time of the field in the resonator are obtained. Moreover, the modulation of electron beams in SCO is investigated. 
It is shown that SCO can be used not only as a radiation source, but also as the beam modulating system. 
The analysis of the expressions obtained allows one to draw the following conclusions. 
Firstly, the resonator with equal-length sections provides the highest efficiency of the energy transfer from the electron beam to the electromagnetic field and the maximum amplitude of the beam current oscillations. 
Secondly, there is the optimal resonator length that provides minimum rise time of the field amplitude and the most efficient conversion of the electron beam energy into the field energy. 
Thirdly, the efficiency of the generator falls dramatically when $\gamma_0 \gg 1$. 
Fourthly, an increase in the beam density leads to the increase in the energy losses of electrons and the degree of beam modulation.

\ifCLASSOPTIONcaptionsoff
  \newpage
\fi

\begin{IEEEbiographynophoto}{Sergei Anishchenko}
 – PhD (Physics and Mathematics), Researcher at Research Institute for Nuclear Problems.

Research Institute for Nuclear Problems BSU (Bobruyskaya str. 11, 220030, Minsk, Belarus).

E-mail: sanishchenko@inp.bsu.by
\end{IEEEbiographynophoto}
\vskip 0pt plus -1fil

\begin{IEEEbiographynophoto}{Vladimir Baryshevsky}
 – Chief Researcher; Doctor of Sciences (Physics and Mathematics); Professor.

Research Institute for Nuclear Problems BSU (Bobruyskaya str. 11, 220030, Minsk, Belarus).

E-mail: bar@inp.bsu.by
\end{IEEEbiographynophoto}
\vskip 0pt plus -1fil

\begin{IEEEbiographynophoto}{Illia Maroz}
 – Master of Science (Physics and Mathematics); Junior Researcher at Research Institute for Nuclear Problems.

Research Institute for Nuclear Problems BSU (Bobruyskaya str. 11, 220030, Minsk, Belarus).

E-mail: miwa-holod@yandex.ru
\end{IEEEbiographynophoto}
\vskip 0pt plus -1fil

\begin{IEEEbiographynophoto}{Anatoli Rouba}
 – PhD (Physics and Mathematics), Senior Researcher at Research Institute for Nuclear Problems.

Research Institute for Nuclear Problems BSU (Bobruyskaya str. 11, 220030, Minsk, Belarus).

E-mail: rouba@inp.bsu.by
\end{IEEEbiographynophoto}
\vskip 0pt plus -1fil

\end{document}